# Airy-beam large volumetric photoacoustic microscopy

Xianlin Song [a, #, *], Ganyu Chen [b, #], Lingfang Song [c, #], Jianshuang Wei [d, e]

[a] School of Information Engineering, Nanchang University, Nanchang 330031, China;
[b] Ji luan Academy, Nanchang University, Nanchang 330031, China;
[c] Nanchang Normal University, Nanchang 330031, China;
[d] Britton Chance Center for Biomedical Photonics, Wuhan National Laboratory for Optoelectronics-Huazhong University of Science and Technology, Wuhan 430074, China;
[e] Moe Key Laboratory of Biomedical Photonics of Ministry of Education, Department of Biomedical Engineering, Huazhong University of Science and Technology, Wuhan 430074, China;
[#] equally contributed to this work

## ABSTRACT

As an emerging nondestructive imaging technology recently, Photoacoustic imaging (PAI), which is based on Photoacoustic effect, combines the advantages: the high resolution and contrast of optical imaging and the high penetration depth of acoustic imaging. Thereinto, as a branch of Photoacoustic imaging, Photoacoustic microimaging inherited the advantages of Photoacoustic imaging. The unique focusing mode of Photoacoustic microimaging can meet the requirements of higher resolution in biological imaging, thus, it gained extensive applications in medical science field. However, on account of using high numerical aperture objective lens strongly focus on Gaussian beam, traditional Photoacoustic microimaging system has shallow depth of imaging field, and its transverse resolution and signal-to-noise ratio deteriorate rapidly outside the focal point, limiting the velocity of large volume imaging. Owing to solve these problems, in this paper, we build a simulation platform for Airy beam photoacoustic microscopy based on K-Wave simulation toolbox. This platform uses Airy beam to inspire initial Photoacoustic signal in large volume and K-Wave simulation toolbox to simulate the propagation, recording and reconstruction process of Photoacoustic signal. As Non-diffraction beam, Airy beam features the capacity of large depth of field, thus, its application could reach the requirement of large depth of field imaging of Photoacoustic microscopy system. Measuring the performances of the constructed Photoacoustic microscopy system, we constructed three-dimensional imaging of the blood vessel. By simulating A-Scan, B-Scan and C-Scan, we measured the performances of this system, such as axial resolution, transverse resolution and depth of field. Meanwhile, the three-dimensional imaging of the vertically tilted fiber also verified the three-dimensional imaging capability of the Airy beam photoacoustic microscopy simulation platform. The establishment of the simulation platform has a significance for the theoretical research of photoacoustic microscopy and its application in biomedicine.

## 1. INTRODUCTION

As an emerging technology recently, Photoacoustic imaging technology, which is based on the Photoacoustic effect and combines optical imaging and acoustic imaging, has been widely used in medical field. Referring to Photoacoustic effect, it means that when the biological tissue is irradiated by pulsed laser, the absorber within the tissue absorbs the light energy, and the local heating causes the tissue to expand, thus generating ultrasonic waves, which could be converted into photoacoustic signals through ultrasonic transducers[1]. A. G. Bell et al. first discovered the Photoacoustic effect in 1880s and proposed that. The same experimental phenomena were firstly observed from solid samples and later also observed from gas and liquid sample[2,9]. In the early 1970s, the rediscovery of Photoacoustic effects in non-gaseous medium made it possible for the Photoacoustic effects to be used to convert information about local radiation absorption into measurable acoustic signals[3]. At about the same period, Photoacoustic effects could be widely used in scanning and detecting gaseous, liquid and solid materials, and it was applied in biomedical imaging in the early 1990s[4,10]. At the same time, with the improvement of Solid photoacoustic theory and the development of laser and ultrasonic detection technologies, the Photoacoustic effect made great breakthroughs in the scattering media and biological tissues field: Photoacoustic imaging technology were proposed. At present, the modes of Photoacoustic imaging technology mainly include the following aspects: Photoacoustic tomography, Photoacoustic microscopy, Photoacoustic endoscopy, Photoacoustic contrast agent and molecular imaging. The photoacoustic tomography mainly uses the non-focused pulse laser as the irradiation source, of which the light scattering makes the inner part of the tissue uniformly irradiated. Thus, with the help of time difference between the acoustic signal of different depth structure before reaching the transducer surface, the Photoacoustic signals of different tomography can be obtained. The mainly-used Photoacoustic imaging technology in this paper is Photoacoustic microimaging. This technology could obtain image point by point by type A and Type B scanning, without relying on complex reconstruction algorithms[5], so the imaging resolution and efficiency are greatly improved, and this technology has been widely concerned in the field of medical imaging. However, on account of using high numerical aperture objective lens strongly focus on Gaussian beam, traditional Photoacoustic microimaging system has shallow depth of imaging field, and its transverse resolution and signal-to-noise ratio deteriorate rapidly outside the focal point, limiting the velocity of large volume imaging. Meanwhile, the Photoacoustic microscopy imaging system has high requirements of the equipment for system construction, which makes the research cost too much. In order to solve problems above, in this paper, a simulation platform of Airy beam Photoacoustic microscopy imaging system based on the K-Wave simulation toolbox is built. The Airy beam with large depth of field and no diffraction can realize the large depth of field imaging of the photoacoustic microimaging system.

## 2. METHOD

In this article, we used Airy beam to illuminate sample, produce the initial pressure field, and K-Wave simulation toolbox was used for acoustic signal transmission and recording while simulate the process of signal reconstruction and reconstruct image, in the set environment, we obtain the three-dimensional image under the condition of two-dimensional distributed Airy beam by point-by-point scanning. By comparing the three-dimensional images with the samples, it is proved that airy beam photoacoustic microscopy can be realized.

**2.1 Airy Beam**

Airy beam is a new non-diffracting beam. In addition to its non-diffraction characteristic, Airy beam has its own characteristic: self-healing and self-bending[6]. In 1979, Berry and Balazs solved the Schrodinger equation under the condition of quantum mechanics, and obtained the solution of the non-diffraction equation under the condition of one-dimensional paraxial, also named Airy wave packet[6].

**2.1.1 Airy Equation**

First, the Schrodinger equation under one-dimensional paraxial conditions is known:

$$i\frac{\partial \varphi}{\partial t} + \frac{\hbar}{2m}\frac{\partial^2 \varphi}{\partial s^2} = 0. \tag{1}$$

In this equation, $\varphi$ said electric field, $\hbar$ said reduced Planck constant, $s$ said dimensionless coordinates.

By deducing equation(1), the following particular solution can be obtained[6]:

$$\varphi = Ai(\frac{Bx}{\hbar^{\frac{2}{3}}}) \tag{2}$$

In the field of optics, the beam propagation under paraxial conditions follows the following diffraction equation(3):

$$i\frac{\partial \varphi}{\partial \xi} + \frac{1}{2k}\frac{\partial^2 \varphi}{\partial s^2} = 0. \tag{3}$$

It can be seen that the Schrodinger one-dimensional linear equation(1) and the diffraction equation under the condition of paraxial have similar forms, thus the solution of Airy function (equation(4)) can be obtained [7]:

$$\varphi(\xi,s) = Ai[s-(\frac{\xi}{2})^2]\exp[i(\frac{s\xi}{2})-i(\frac{\xi}{12})^3] \qquad (4)$$

In equation(4), $\exp(as)$ represents exponential decay, $a$ is defined as the limited energy to constraint the airy beam energy, make the airy beam can be implemented in reality.

Take the Fourier transform to $\varphi(\xi=0,s)=Ai(s)\exp(as)$, mathematical expression of Kode Airy beam in K space could be get[8]:

$$\varphi(k) = \exp(-ak^2)\exp[\frac{i}{3}(k^3-3a^2k-ia^3)] \qquad (5)$$

Through equation(5), we could obtain the one-dimensional Airy beam equation under the condition of finite energy:

$$\varphi(\xi,s) = Ai[s-(\frac{\xi}{2})^2+ia\xi]\exp[as-(\frac{a\xi^2}{2})-(\frac{\xi}{12})^3+i(\frac{a^2\xi}{2})+i(\frac{s\xi}{2})]. \qquad (6)$$

The two-dimensional airy beam equation is:

$$\begin{aligned}\varphi(\xi,s) &= Ai[s_x-(\frac{\xi}{2})^2+ia\xi]\exp[as_x-(\frac{a\xi^2}{2})-(\frac{\xi}{12})^3+i(\frac{a^2\xi}{2})+i(\frac{s_x\xi}{2})] \\ &\times Ai[s_y-(\frac{\xi}{2})^2+ia\xi]\exp[as_y-(\frac{a\xi^2}{2})-(\frac{\xi}{12})^3+i(\frac{a^2\xi}{2})+i(\frac{s_y\xi}{2})]\end{aligned} \qquad (7)$$

### 2.1.2 Generation of Airy Beam

It can be seen from the formula that the light intensity diagram of Airy beam is a spot with a 90° side lobe on both sides, the main lobe in the center has the strongest light intensity, while the light intensity of the side lobe decreases according to the distance. When the Airy beam propagates along the Z axis, its trajectory is a quadratic function.

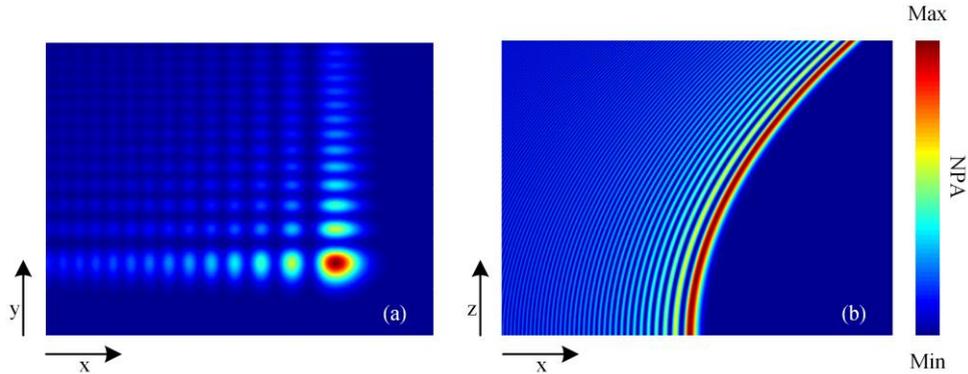

Figure 1. (a) (b) is the spot formed by Airy' s equation. (a) is the light spot figure of Airy Beam formed in the x-y plane. (b) is the light spot figure of Airy Beam formed in the x-z plane.

From Figure 1, It can be seen that airy beam spot is 90° distributed in the side lobe, with the strongest light intensity in the central main lobe. And propagation of Airy Beam along the quadratic function, showing its self-bending properties.

## 2.2 Set up K-wave simulation environment

### 2.2.1 K-wave

In this paper, the k-Wave toolbox is used to simulate the photoacoustic signals. The K-Wave simulation toolbox was developed by Bradley Treeby, Ben Cox and Jiri Jaros in 2009, based on MATLAB[11]. In the K-Wave simulation toolbox, we use the function KspaceFirstOrder3D (Kgrid, medium, source, sensor), where kgrid means the grid, medium means the propagation medium (usually water), source means the sound source, and sensor represents the information of the detector. This function according to the set step length, calculates the sound pressure value of the photoacoustic signal propagating in the medium, and records it as the photoacoustic signal. The function kgrid = makeGrid (Nx, Ny, Nz, dx, dy, dz) is used to set the 3D grid, in where Nx represents the number of grids in the x direction, dz represents the size of grids, and the same for y and z direction.

When conducting a simulation, we followed these steps:

 (1) Set the structure parameters, such as optical wavelength, optical focus position, scanning precision parameters, dimensionless coordinate scale, attenuation factor of finite energy Airy beam, sound velocity in medium, mesh number, mesh size. (2) Calculate the acoustic pressure field of the simulated photoacoustic signal. (3) Record and analyze the photoacoustic signals.

### 2.2.2 System Parameters

In this paper, the tissue sample parameters of photoacoustic microimaging were established as Table 1 below:

Tabel1. Data of each item for basing K-wave simulation environment.

| Item | Data |
|---|---|
| Wavelength (m) | $532 \times 10^{-9}$ |
| Velocity of sound in water (m/s) | 1500 |
| Zfocus (m) | $1.5 \times 10^{-4}$ |
| Number of Kgrid | $60 \times 60 \times 60$ |
| Size of Kgrid (m) | $5 \times 10^{-6}$ |
| Transverse accuracy | 2 |
| Scale(m) | $3 \times 10^{-6}$ |

# 3. RESULT

## 3.1 Scanning path

In this paper, we combined A-Scan, B-Scan and C-Scan to simulate. Shown as figure 2, from the starting point, the light spot moves along the X-axis, scanning an entire row before moving to the next column, aiming to scanning the entire area point by point.

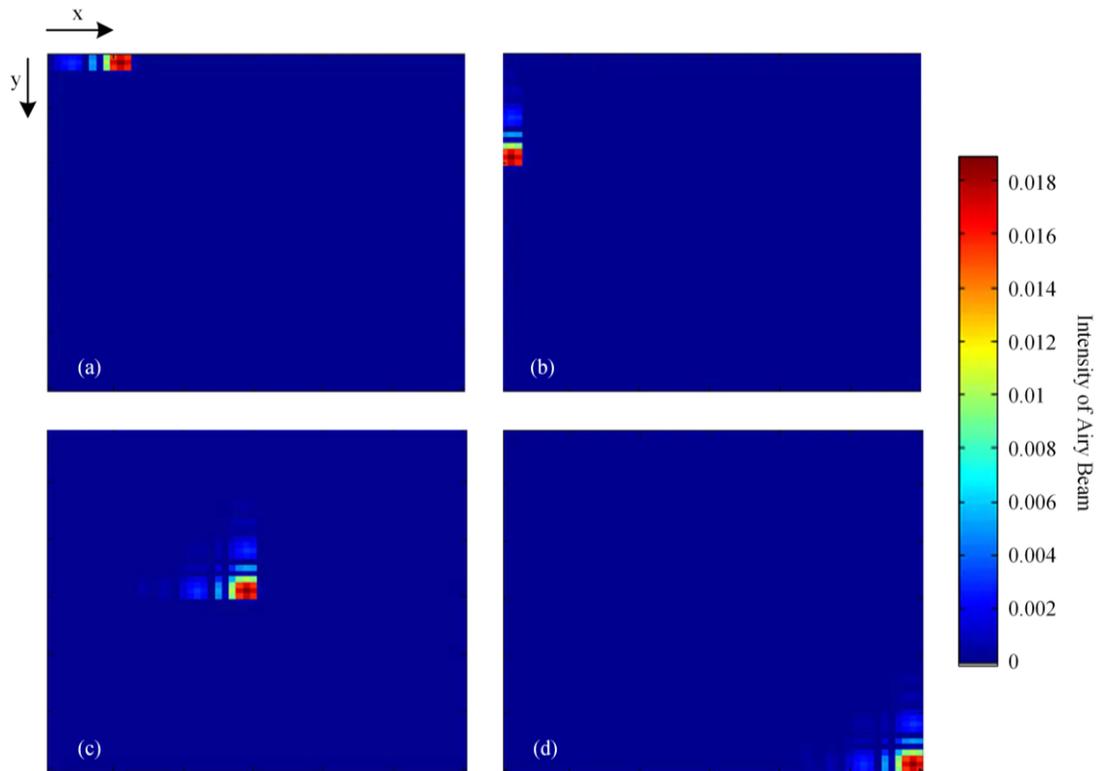

Figure 2. The spot moved from (a) to (c). To entirely scan all of the point, we used A-scan in x direction while B-scan in y direction.

### 3.2 Simulation result

During the scanning, the photoacoustic signal obtained from the scanning sample is convolved with the point spread function obtained from each scanning point to obtain the final photoacoustic microimaging image, which is the final photoacoustic image. As shown in Figure 3, Figure 3 (a) is the sample. Figure 3 (b) is the result of A-scan, showing development of light intensity along with time. Figure 3 (c) is the result of B-scan. Figure 3 (d) is the result of D-scan, also the final result of this simulation. It can be seen that there is a strong sidelobe photoacoustic signal around the obtained blood vessel image, which is determined by the characteristics of the Airy beam spot. The sidelobe of the Airy beam will affect the resolution of the image, and then make the image becomes more vague.

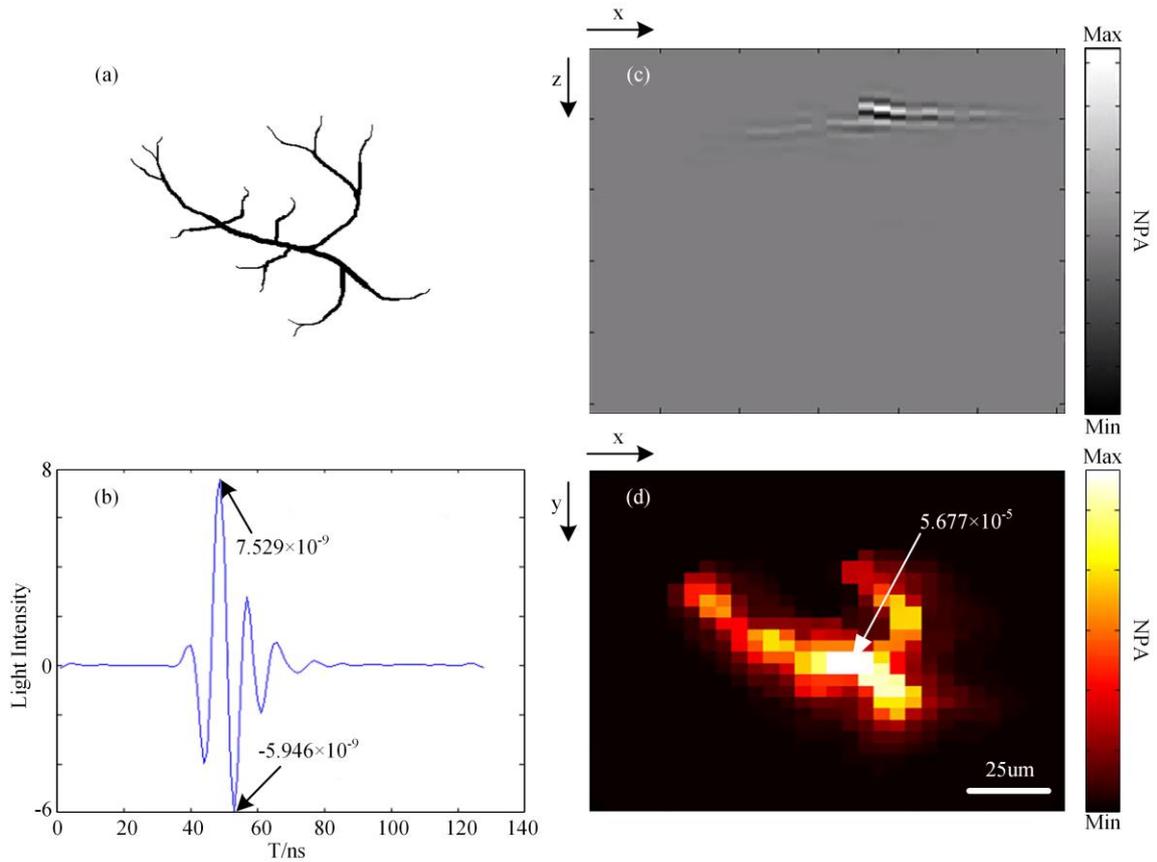

Figure 3. The result of scanning. (a) is the sample. (b) is the result of A-scan, showing development of light intensity along with time. (c) is the result of B-scan. (d) is the result of D-scan, also the final result of this simulation.

## 4. CONCLUSION

We build a simulation platform for Airy beam photoacoustic microscopy based on K-Wave simulation toolbox in MATLAB. The sample was located in the middle of z direction in the environment. The propagation of Airy beam irradiated on the sample was simulated by solving the Schrodinger equation. And the intensity of Airy beam was simulated by Airy wave pocket. The simulation result showed that: in A-scan, the intensity of light waved over time, reached peak at 49ns; in C-scan, the image agrees with the sample. This research showed that the Airy beam with large depth of field and no diffraction can realize the large depth of field imaging of the photoacoustic microimaging system. The establishment of the simulation platform has a significance for the theoretical research of photoacoustic microscopy and its application in biomedicine.